# Entangled quantum Otto heat engines based on two-spin systems with the Dzyaloshinski–Moriya interaction

Li-Mei Zhao(赵丽梅), Guo-Feng Zhang(张国锋)[*]

*Key Laboratory of Micro-Nano Measurement-Manipulation and Physics (Ministry of Education), School of Physics and Nuclear Energy Engineering, Beihang University, Xueyuan Road No. 37, Beijing 100191, China*

**Abstract:** We construct an entangled quantum Otto engine based on spin-1/2 systems undergoing Dzyaloshinski–Moriya (DM) interaction within a varying magnetic field. We investigate the influence of the DM interaction on basic thermodynamic quantities, including heat transfer, work done, and efficiency and find that the DM interaction importantly influences the engine's thermodynamics. We obtain an expression for engine efficiency, finding it to yield the same efficiency for antiferromagnetic and ferromagnetic coupling. A new upper bound, nontrivially consistent with the second law of thermodynamics, is derived for engine efficiency in the case of non-zero DM interaction.

**Keywords:** Quantum heat engine; Spin-1/2 systems; Dzyaloshinski–Moriya interaction (DM)

PACS numbers: 03.67.Hk; 03.65.Ud; 75.10.Jm

## I. INTRODUCTION

With the rapid development of quantum information, the relation between quantum mechanics and thermodynamics has become an active area of research [1]. The quantum heat engines (QHEs) was first constructed by Kieu which is a two-level quantum system with the basic thermodynamic quantities, i.e., heat transferred, net work done in a cycle and efficiency [2]. Since then, a significant amount of effort has been devoted to studies of all kinds of QHEs, such as coupled spins system [3–25], harmonic oscillator system [26-29], two level or multilevel system [30-31], and cavity quantum electrodynamics system [32,33].

Recently, the physics of semiconductors with a spin-orbit interaction has attracted a lot of attention [34]. The Dzyaloshinskii-Moriya (DM) interaction [35, 36] which is arising from the spin-orbit coupling can generate many surprising characteristics. Then the influence of the DM interaction on the performance of the basic thermodynamical quantities must be considered.

---

Actually, many investigations have been carried out to explore various possible improvement on the quantum coupling systems with DM interaction by virtue of rapidly developing quantum mechanics, and many surprising properties are reported [37-43]. For example, Zhang extended Kieu's work by considering the quantum entanglement heat engine with DM interaction [39]. Wang et al. research the thermal entanglement in the two-qubit anisotropic XXZ model and the Heisenberg model with DM interactions [43].

However, none of the QHEs mentioned above involving the specific effects of spin-orbit coupling with the performance of the quantum Otto heat engines. Moreover, local thermodynamics was never taken on account in the previous studies. It is a well-known fact that the dynamics of subsystem may be very different from that of the whole system for an entangled case. To enrich research in QHEs, the investigation about the influence of DM interaction on its thermodynamic characteristics, especially on the local thermodynamics, should be included. The paper will show two cases (spin-spin coupled system with variable DM interaction in a fixed or a changeable magnetic field) in which thermodynamic characteristics are given numerically and analytically. From practical and experimental points of view, the two models are also easily realized [44-45].

The outline of the paper is as follows. In Sec. II, we describe our working substance and the model for QHEs. In Sec. III, the positive work condition (PWC) and engine's efficiency is discussed for the case with different DM interaction for the fixed spin exchange constant and magnetic field. In Sec. IV, another case of different external magnetic field is analyzed from the perspective of local work and local efficiency. Result shows that DM interaction had profound effects on the performance of the global or local quantum heat engine for both cases. More important, a new upper bound is derived for the efficiency of the global cycle within a certain range of parameter values. The final section is devoted to the conclusion of our main findings.

## II. THE MODEL AND SOLUTIONS

We consider two spin-1/2 particles as working substance of the system. The particles are coupled by an DM interaction in an external magnetic field $B$, the Hamiltonian [39] is given by

$$H_{DM} = \frac{J}{2}\left[(\sigma_{1X}\sigma_{2X} + \sigma_{1Y}\sigma_{2Y}) + \vec{D} \cdot (\vec{\sigma_1} \times \vec{\sigma_2})\right] + B(\sigma_Z^1 + \sigma_Z^2), \tag{1}$$

where $D$ is the DM vector coupling and $J$ is the exchange coupling constant. While $J > 0$ and $J < 0$ correspond to the antiferromagnetic and the ferromagnetic cases, respectively. For simplicity, we choose $\vec{D} = D\vec{z}$, then the Hamiltonian becomes

$$H_{DM} = J[(1 + iD)\sigma_{1+}\sigma_{2-} + (1 - iD)\sigma_{1-}\sigma_{2+}] + B(\sigma_Z^1 + \sigma_Z^2). \tag{2}$$

The four eigenvectors of Eq. (2) are

$$|\psi_1> = |00>,$$

$$|\psi_2> = |11>,$$

$$|\psi_3> = \frac{1}{\sqrt{2}}(|01> + e^{-i\theta}|10>),$$

$$|\psi_4> = \frac{1}{\sqrt{2}}(|01> - e^{-i\theta}|10>). \tag{3}$$

And the corresponding eigenvalues can be explicitly expressed as $E_1 = -2B, E_2 = 2B, E_3 = J\sqrt{1 + D^2} = -E_4$, and $\theta = \arctan D$. For this physical system, the density operator when the system is at thermal equilibrium can be written as $\rho = \sum_i p_i |\psi_i><\psi_i|$ with $p_i = \exp(-\beta H_{DM})/Z$. The partition function of the system is described by $Z = \text{Tr}[\exp(-\beta H_{DM})]$, where $\beta = 1/(k_B T)$ and $k_B$ is the Boltzmann constant, $T$ is the temperature. We set Boltzmann's constant $k_B = 1$.

## III. BASIC THERMLDYNAMIC CHARACTERISTICS FOR DIFFERENT DM PARAMETERS

The expectation value of the the energy of the system with four coupling discrete energy levels is: $U = <H> = \sum_i p_i E_i$, where $E_i$ are the energy levels of the system and $p_i$ are the corresponding occupation probabilities. Infinitesimally, we have $dU = \sum_i (p_i dE_i + E_i dp_i)$. With the first law of thermodynamics, the infinitesimal heat transferred can identified as $dQ = \sum_i E_i dp_i$ and the net work done is $dW = \sum_i p_i dE_i$.

It is supposed that the external magnetic field is kept constant throughout and only DM interaction is varied during the cycle. The cycle of the entangled quantum heat engine consists of two isochoric and two adiabatic processes. The four quantum thermodynamic stages are described below:

Stage 1: The working medium is in thermal equilibrium with the external magnetic field $B$ is coupled to a hot bath of temperature $T_1$ which keep with $D = D_1$. After some time, the occupation probabilities of each eigenstate changes from $p_i$ to $p_i'$, respectively. while each eigenenergy is kept fixed at the value $E_i$. In this process, only heat is transferred and no work done.

Stage 2: the system is isolated from the hot reservoir and undergoes an adiabatic process which the DM interaction is changed from $D_1$ to $D_2$ such that $D_1 < D_2$. During this process, the quantum adiabatic theorem is assumed to hold [14], according to which the rate of the process enough to maintain the individual occupation probability of each energy level. Each eigenenergy is varied from $E_i$ to the smaller value $E'_i$. Work is performed by the system during this step.

Stage 3: the system is next brought into thermal contact with a cold bath of temperature $T_2 (< T_1)$ with its energy structure is kept fixed. After the irreversible thermalization process, the occupation probability of each eigenstate changes from $p'_i$ (i=1,2,3,4) to $p_i$. Only heat is transferred in the stage.

Stage 4: the system is detached from the bath and the DM interaction from $D_2$ is increased from $D_1$ but kept the occupation probability $p_i$ fixed. In this process, the energy structure is varied from $E'_i$ to $E_i$. Thus, an amount of work is performed by the system.

Finally the system is attached to the hot reservoir again. Thus the system returns to the initial state, completing the heat cycle. After a simple calculation, the heat transferred in Stage 1 is:

$$Q_H = \sum_i E_i (p_i - p'_i) = J\sqrt{D_1^2 + 1}\,(p_3 - p'_3 + p'_1 - p_1) + 2B(p_4 - p'_4 + p'_2 - p_2). \qquad (4)$$

The heat transferred in Stage 3 is:

$$Q_L = \sum_i E'_i (p'_i - p_i) = -J\sqrt{D_2^2 + 1}\,(p_3 - p'_3 + p'_1 - p_1) - 2B(p_4 - p'_4 + p'_2 - p_2). \qquad (5)$$

In the above, $Q_H$ and $Q_L$ correspond to absorbption and release of heat from and to the heat baths, respectively. The net work done in Stage 2 and Stage 4 per cycle is:

$$W = Q_H + Q_L = J\left(\sqrt{D_1^2 + 1} - \sqrt{D_2^2 + 1}\right)(p_3 - p'_3 + p'_1 - p_1). \qquad (6)$$

Note that $W > 0$ corresponds to work performed by the system. The efficiency of the engine cycle can be defined as:

$$\eta = \frac{W}{Q_H} = \frac{J\left(\sqrt{D_1^2+1} - \sqrt{D_2^2+1}\right)(p_3 - p'_3 + p'_1 - p_1)}{J\sqrt{D_1^2+1}(p_3 - p'_3 + p'_1 - p_1) + 2B(p_4 - p'_4 + p'_2 - p_2)}. \qquad (7)$$

By numerical calculation we can plot the isoline maps of the variation of basic thermodynamic quantities with $D_1$ and $D_2$ in two given representative values $B = 4$ and $B = 6$, as shown in Figs. 1–3. It is shows that the DM interaction have an important effect on the properties of basic thermodynamic quantities. We also find the condition which the net work and efficiency are positive is $D_1 < D_2$. From the Eq. (6), if $D_1 < D_2$, in order to make $W > 0$, we must have $(p_3 - p'_3 + p'_1 - p_1) < 0$, which means $B/T_1 < B/T_2$. Moreover, from the comparison of Figs.1 and Figs.3, there are the same net work and efficiency for both $J > 0$ and $J < 0$. This result can be easily proved from the calculation of Eq.(6) and (7). Finally, $Q_H > -Q_L > 0$ is always true as long as condition $W > 0$ is satisfied. And the maximum efficiency is smaller than the Carnot efficiency $\eta_c = 1 - T_2/T_1 = 0.5$. Therefore, the model case does not violate the second law of thermodynamic.

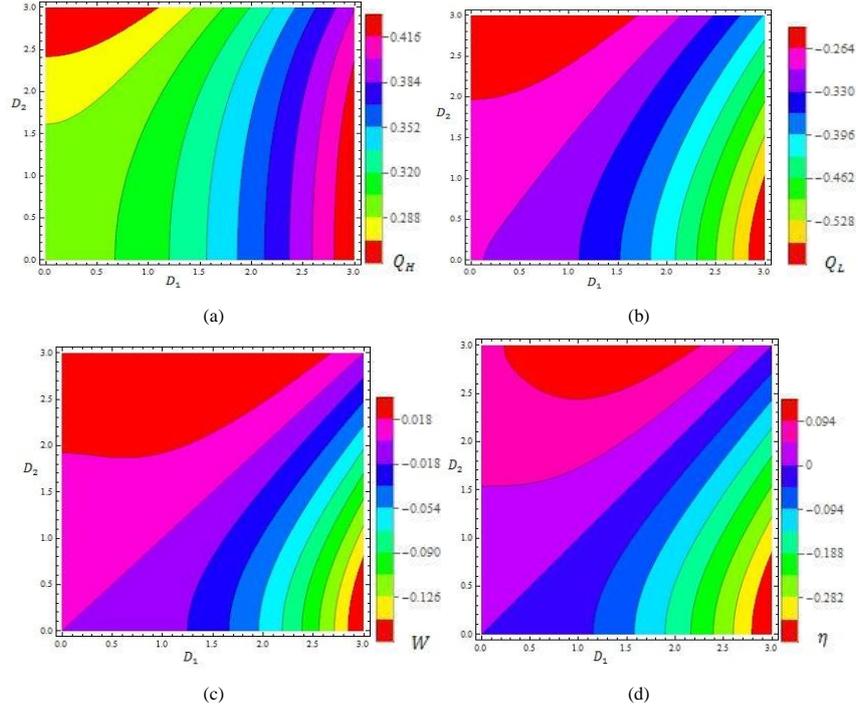

FIG.1. Variations of heat transferred (a) $Q_H$, (b) $Q_L$, (c) net work input $W$ and (d) the efficiency $\eta$ of the QHE with variables $D_1$ and $D_2$ in isoline map for parameters $T_1 = 2, T_2 = 1, J = 1$ and $B = 4$.

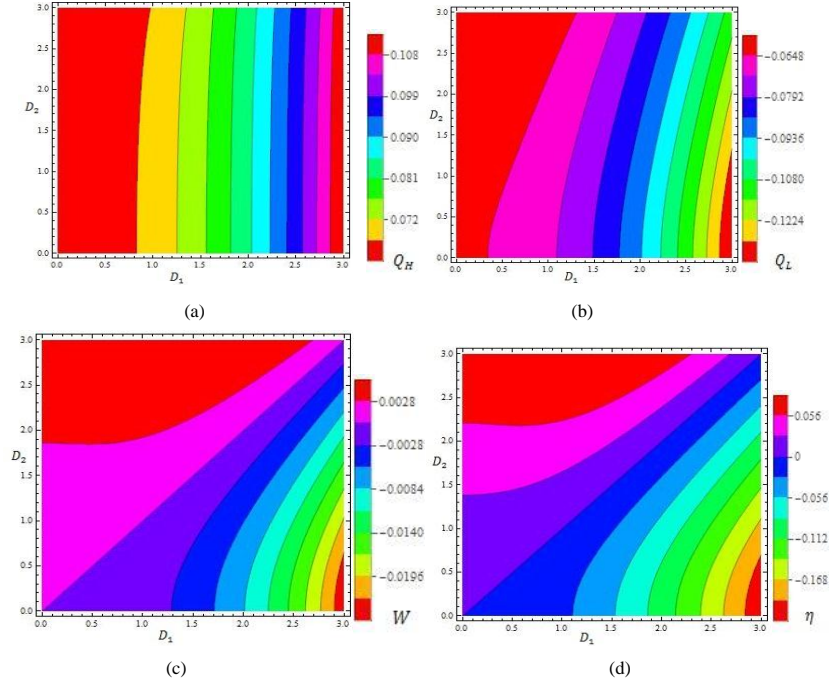

FIG.2. Variations of heat transferred (a) $Q_H$, (b) $Q_L$, (c) net work input $W$ and (d) the efficiency $\eta$ of the QHE with variables $D_1$ and $D_2$ in isoline map for parameters $T_1 = 2, T_2 = 1, J = 1$ and $B = 6$.

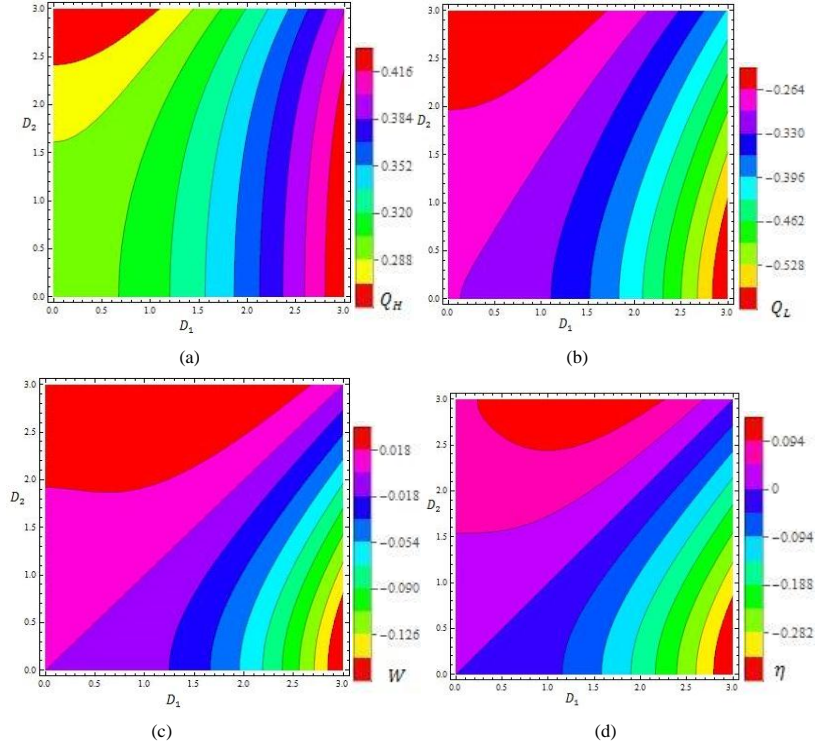

FIG.3. Variations of heat transferred (a) $Q_H$, (b) $Q_L$, (c) net work input $W$ and (d) the efficiency $\eta$ of the QHE with variables $D_1$ and $D_2$ in isoline map for parameters $T_1 = 2, T_2 = 1, J = -1$ and $B = 4$.

## IV. THE EFFECT OF DIFFERENT MAGNETIC FIELD ON THE THERMODYNAMIC CHARACTERISTICS FOR THE SAME DM INTERACTION

In this part, the effect of different magnetic field on bascis thermodanamics characteristics will be studied in detail. Local thermodynamic process for the individual spin is our key point. From practical and experimental points of view, it is easier to change the external magnetic field $B$. The four stages in quantum Otto cycle are described as follows:

Stage 1: the system is coupled to a hot bath of temperature $T_1$ with the external magnetic field $B = B_1$. After some contact time, each eigenenergy is kept fixed at the value $E_i$ while the occupation probabilities of each eigenstate changes from $p_i$ to $p_i'$.

Stage 2: then the system is isolated from the hot bath. This is an adiabatic process which changes magnetic field from $B_1$ to $B_2$. According to quantum adiabatic theorem [14], the compression rate is sufficiently slow enough to keep $p_i$ ($i$=1,2,3,4) fixed. In this process, each eigenenergy is varied from $E_i$ to the smaller value $E_i'$.

Stage 3: the system is next brought into thermal contact with a cold bath at temperature $T_2$ ($T_1 > T_2$) for some time. After the irreversible thermalization process, the occupation probabilities $p_i'$ back to $p_i$ corresponding to the thermal state with $B = B_2$.

Stage 4: the system is removed from the bath and undergoes a quantum adiabatic process with the magnetic field changes from $B_2$ back to $B_1$. In this process, the occupation probability $p_i$ is kept fixed, but the energy structure is varied from $E_i'$ to $E_i$.

After a simple calculation, we can get the heat transferred in Stage 1 is

$$Q_H = \sum_i E_i(p_i - p_i') = J\sqrt{D^2 + 1}\,(p_3 - p_3' + p_1' - p_1) + 2B_1(p_4 - p_4' + p_2' - p_2). \tag{8}$$

The heat transferred in Stage 3 of this cycle is

$$Q_L = \sum_i E_i'(p_i' - p_i) = -J\sqrt{D^2 + 1}\,(p_3 - p_3' + p_1' - p_1) - 2B_2(p_4 - p_4' + p_2' - p_2). \tag{9}$$

In the above, $Q_H$ and $Q_L$ correspond to absorbption and release of heat from and to the heat baths, respectively. The net work done per cycle is

$$W = 2(B_1 - B_2)(p_4 - p_4' + p_2' - p_2). \tag{10}$$

Note that $W > 0$ corresponds to work performed by the system. The efficiency of the QHE is

$$\eta = \frac{W}{Q_H} = \frac{2(B_1 - B_2)(p_4 - p_4' + p_2' - p_2)}{J\sqrt{D^2+1}(p_3 - p_3' + p_1' - p_1) + 2B_1(p_4 - p_4' + p_2' - p_2)}. \tag{11}$$

In the following, we discuss the individual spins thermodynamic process in the cycle [5]. Let $\rho_{12}$ and $\rho_{12}'$ represent the thermal equilibrium states when the system in Stage 1 and Stage 3 respectively. The density matrices are given by

$$\rho_{12} = \begin{pmatrix} p_2 & 0 & 0 & 0 \\ 0 & \frac{p_1+p_3}{2} & \frac{p_3-p_1}{2} & 0 \\ 0 & \frac{p_3-p_1}{2} & \frac{p_1+p_3}{2} & 0 \\ 0 & 0 & 0 & p_4 \end{pmatrix},$$

$$\rho_{12}' = \begin{pmatrix} p_2' & 0 & 0 & 0 \\ 0 & \frac{p_1'+p_3'}{2} & \frac{p_3'-p_1'}{2} & 0 \\ 0 & \frac{p_3'-p_1'}{2} & \frac{p_1'+p_3'}{2} & 0 \\ 0 & 0 & 0 & p_4' \end{pmatrix}. \tag{12}$$

In Stage 1, Let $\rho_1$ and $\rho_2$ be the reduced density matrices for the first and the second spin, respectively. Let $\rho_1'$ and $\rho_2'$ be the reduced density matrices in Stage 3. Then from the normalization constraints, $\sum_n p_n = \sum_n p_n' = 1$, we get the reduced density matrices

$$\rho_1 = \rho_2 = \begin{pmatrix} \frac{1}{2} - (\frac{p_2-p_4}{2}) & 0 \\ 0 & \frac{1}{2} + (\frac{p_2-p_4}{2}) \end{pmatrix},$$

$$\rho_1' = \rho_2' = \begin{pmatrix} \frac{1}{2} - (\frac{p_2'-p_4'}{2}) & 0 \\ 0 & \frac{1}{2} + (\frac{p_2'-p_4'}{2}) \end{pmatrix}. \tag{13}$$

Let $H_l$ and $H'_l$ be the local Hamiltonians for individual spins with eigenvalues $(B_1, -B_1)$ and $(B_2, -B_2)$ in Stage 1 and Stage 3 respectively. Then the local heat transmission is defined as $q_1 = \text{Tr}[(\rho_1 - \rho'_1)H_l]$ and $q_2 = \text{Tr}[(\rho_2 - \rho'_2)H'_l]$. The explicit expressions can be described by

$$q_1 = B_1(p_4 - p'_4 + p'_2 - p_2),$$
$$q_2 = -B_2(p_4 - p'_4 + p'_2 - p_2). \quad (14)$$

We get the global heat absorbed by the system and the heat released to the second bath can be written as

$$Q_H = J\sqrt{D^2 + 1}(p_3 - p'_3 + p'_1 - p_1) + 2q_1,$$
$$Q_L = -J\sqrt{D^2 + 1}(p_3 - p'_3 + p'_1 - p_1) + 2q_2. \quad (15)$$

Further, the net work done by an individual spin is

$$w = q_1 + q_2 = (B_1 - B_2)(p_4 - p'_4 + p'_2 - p_2). \quad (16)$$

Compare (10) with (16), the relationship about total work and local work is

$$W = 2w. \quad (17)$$

The total work are the sum of two local efforts. The efficiency of the local system can be defined as $\eta_0 = w/q_1 = 1 - B_2/B_1 = 0.25$

In the following, we make a detailed analysis of two cases under the impact of the external magnetic field on this cycle system.

**(A) The case $B_1 > B_2$**

By fixing some parameters we can know the roles of the other parameters and the variation of the DM intercation. Then the thermodynamic quantities are plotted versus the DM intercation $D$ value in Fig. 4., for $B_1 = 8, B_2 = 6, J = 1$ and $T_1 = 2, T_2 = 1$. It is intuitively find that the DM intercation $D$ value not only affects the shape of the heat transferred, the net work but also affects its value. With the increase of the constant $D$, the heat engine cycle turn into a refrigeration cycle. Here we are interested in the situation that the net work is positive.

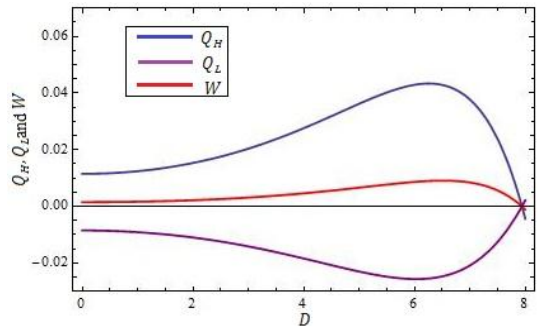

FIG. 4. The heat exchanged $Q_H$, $Q_L$ and net work input $W$ are plotted vs. the interaction $D$, for $B_1 = 8, B_2 = 6, J = 1$ and $T_1 = 2, T_2 = 1$.

In order to $W > 0$, from Eq. (10), the inequality can write $p_2 - p_4 < p_2' - p_4'$. And from Eq. (8) $Q_H > 0$, we have one of the following two possibilities: (i) $p_3 - p_3' + p_1' - p_1 > 0$ or (ii) $p_3 - p_3' + p_1' - p_1 < 0$. Along with the possibility (ii), we get the inequality $(p_4 - p_4' + p_2' - p_2) > -(J\sqrt{D^2+1}/2B_1)(p_3 - p_3' + p_1' - p_1)$. The Eq. (8) become

$$-J\sqrt{D^2+1}(p_3 - p_3' + p_1' - p_1) = Q_H\left(1 - \frac{B_1\eta}{B_1 - B_2}\right). \tag{18}$$

or $J\sqrt{D^2+1}(p_1 - p_1' + p_3' - p_3) = Q_H(1 - \eta/\eta_0)$, where $\eta = W/Q_H$ is the efficiency of the global coupled engine and $\eta_0 = 1 - B_2/B_1 = 0.25$ is the efficiency of the local engine.

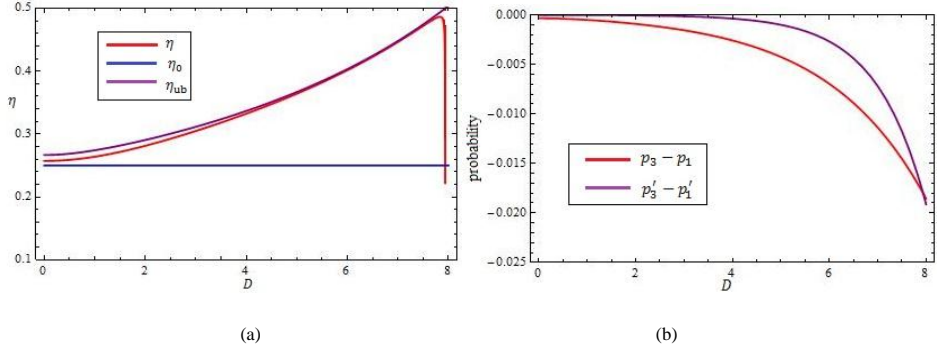

(a) (b)

FIG. 5. The efficiency(a) and the probabilities (b) versus the interaction $D$, for values $B_1 = 8, B_2 = 6, J = 1$ and $T_1 = 2, T_2 = 1$

From Fig.5, it is observed that $\eta$ as function of $D$ shows a non-monotonic behavior. We can see a region in Fig.5 (b), $p_3 - p_1 < p_3' - p_1'$, which efficiency of the global engine $\eta$ can be higher than the efficiency of the local case $\eta_0$. As $D$ up to a certain value of $D_m$, we have $p_3 - p_1 > p_3' - p_1'$. Due to Eq. (18), the term $J\sqrt{D^2+1}(p_3 - p_3' + p_1' - p_1)$ is positive which makes the efficiency $\eta$ less than $\eta_0$. For the case $\eta = \eta_0$, we constructed a function $F(D_m) = p_3 - p_1 - p_3' + p_1' = 0$, which implies

$$D_m = 4B_1^2/J^2 - 1. \tag{19}$$

Thus, the condition of the efficiency $\eta > \eta_0$ is $D < 4B_1^2/J^2 - 1$.

Within the general discussion, the efficiency of the global engine $\eta$ can be higher than the efficiency of the local case $\eta_0$ when $p_3 - p_1 < p_3' - p_1'$. Therefore, this condition is equal to $(p_2 - p_4)/(p_1 - p_3) < (p_2' - p_4')/(p_1' - p_3')$. From the expressions for the probabilities, the inequality can be simplified to write $B_1/T_1 < B_2/T_2$. Then we obtain $p_4 > p_4'$ and $p_3 > p_3'$. The normalisation of the probabilities gives $p_1 + p_2 < p_1' + p_2'$. Therefore, $p_3 - p_3' + p_1' - p_1 < 0$ and $p_4 - p_4' + p_2' - p_2 > 0$. We can write $p_1 > p_1', p_2 < p_2'$, which means

$$\frac{2B_1 - J\sqrt{D^2+1}}{T_1} < \frac{2B_2 - J\sqrt{D^2+1}}{T_2}. \tag{20}$$

For convenience, we write above inequality as: $-(p_3 - p_1 + p_1' - p_3') < (p_4 - p_2 - p_4' + p_2')$. This ensures that an upper bound exists for the efficiency, given as:

~ 9 ~

$$\eta = \frac{\eta_0}{1-\frac{J\sqrt{(D^2+1)}(p_1-p_1'+p_3'-p_3)}{2B_1(p_4-p_2+p_2'-p_4')}} < \frac{1-\frac{B_2}{B_1}}{1-\frac{J\sqrt{D^2+1}}{2B_1}} = \eta_{\text{ub}}, \qquad (21)$$

where $\eta_{\text{ub}}$ is the upper bound for efficiency when the coupled case is higher than the uncoupled case. It can be shown that the condition $D < 4B_1^2/J^2 - 1$ implies the following relations between the efficiency $\eta > \eta_0$. From the above discussion about the Eq. (20) and (21), we have $\eta_{\text{ub}} < \eta_c$, where $\eta_c = 1 - T_2/T_1 = 0.5$ is the Carnot bound. It can be directly seen that the upper bound $\eta_{\text{ub}}$ is independent of the reservoir temperatures, and is less than the Carnot limit $\eta_c$ within $D < 4B_1^2/J^2 - 1$. From Fig.5, the efficiency of the global engine $\eta$ is higher than the local $\eta_0$ at the value $D = 0$. Moreover, there exits a relationship which the efficiency is $\eta_0 < \eta < \eta_{\text{ub}}$ at the point $D = 0$. The result is different from the system with coupled spin-1/2 particles [G. Thomas and R. S. Johal, Phys. Rev. E 83, 031135 (2011)]. In this regime $D < 4B_1^2/J^2 - 1$, the ordering of energy levels is

$$-2B_1 < -J\sqrt{D^2+1} < J\sqrt{D^2+1} < 2B_1, \qquad (22)$$

and which after the first quantum adiabatic process, becomes

$$-2B_2 < -J\sqrt{D^2+1} < J\sqrt{D^2+1} < 2B_2. \qquad (23)$$

Finally, it can be directly seen that the second law of thermodynamics is reasonable in this case. The same conclusion is consistent for anferromagnetic case.

**(B) The case $B_1 < B_2$**

In this case, the magnetic field is increased from $B_1$ to $B_2$ during the first quantum adiabatic process. If there is no interaction between the spins, the system cannot work as an engine in this case because the condition $W > 0$ will not be satisfied [2]. The conditions $T_1 > T_2$ and $B_1 < B_2$ directly leads to $p_4 > p_4'$, $p_3 > p_3'$. Since $W > 0$, the Eq. (10) can be written as $p_2 > p_2'$. From the normalisation of probabilities, we have $p_1 < p_1'$.

According to Eqs. (14), the local work $q_1 < 0$ and $q_2 > 0$. This means the local heat is absorbed from the cold bath and given to the hot bath. But globally we do have $Q_H > 0$ and $Q_L < 0$. Then the heat transferred direction of the global cycle system and the local system are the opposite.

The two-spin engine is consistent with the second law of thermodynamics, although locally it is seem to have a violation. The efficiency in the local system is $\eta_0 = w/q_2 = 1 - B_1/B_2 = 0.25$. This special phenomenon has resolved to use the concept of local effective temperatures in Ref. [5].

**V. CONCLUSIONS**

In conclusion, by introducing the quantum interpretations of an entangled quantum Otto heat engine with DM coupling interaction, we have broadened the investigations of quantum heat engine. It is found that the the DM interaction leads to enhancement the thermodynamic quantities, i.e., heat transferred, the extracted work and the efficiency. We propose that $D_1 < D_2$ is the sufficient criterion for $W > 0$ for fixed magnetic field and spin-spin exchange coupling. There are the same conclusion expressions of both antiferromagnetic and ferromagnetic coupling. The total work performed is the sum of work obtained from that of the two spins locally for given fixed DM coupling. When $B_1 > B_2$, the condition under which the efficiency of the global engine $\eta$ can be higher than the local case $\eta_0$ is $B_1/T_1 < B_2/T_2$. An upper bound to the efficiency is derived in the regime $D < 4B_1^2/J^2 - 1$, which also implies that the bound itself is limited by Carnot value. For the case $B_1 < B_2$, we find a phenomenon that the heat transfer direction of the global cycle system and the local system are the opposite. The second law of thermodynamics is shown to be valid all the while in the entangled system.

## ACKNOWLEDGEMENT

This work is supported by the National Natural Science Foundation of China (Grant No. 11574022) and the Fundamental Research Funds for the Central Universities of Beihang University (Grant No. YWF-17-BJ-Y-70).